# Outage Probability of Overhearing Amplify-and-Forward Cooperative Relaying

Yu Zhang[1], Ke Xiong[2,3*]

[1] School of Computer and Communication Engineering, University of Science and Technology Beijing, Beijing 100083, China
[2] School of Computer and Information Technology, Beijing Jiaotong University, Beijing 100044, China
[3] National Mobile Communications Research Laboratory, Southeast University, Nanjing 210018, China
*contact.author: kxiong@bjtu.edu.cn

*Abstract*—This paper investigates the outage performance of overhearing amplify-and-forward (AF) cooperative relaying, where a source transmits information to its destination through multiple helping overhearing AF relays with space-time network coding (STNC) employed. Firstly, the transmission protocol of such a relaying system, i.e., cooperative relaying with overhearing AF relays based on STNC (STNC-OHAF) is presented. Then, the instantaneous end-to-end SNR expression of STNC-OHAF is analysed. Based on this, an explicit expression of the outage probability for STNC-OHAF over independent but not necessarily identically distributed (i.n.i.d) Rayleigh fading channels is theoretically derived. Numerical results validate our theoretical analysis and show that by introducing overhearing among relays, the outage performance of the system can be greatly improved. It also shows that there is a trade-off between system sum outage capacity and the transmitted number of symbols.

## I. INTRODUCTION

Cooperative relaying [1] has been widely considered as a promising solution to resolve the problem of deploying multiple-input multiple-output (MIMO) in single-antenna systems while inheriting MIMO benefits in improving reliability. In order to avoid the imperfect synchronization issue [4] when two or more nodes transmits signal at the same time in cooperative relaying systems, time-division multiple access (TDMA) is regarded as the most effective and practical technique [5]. Nevertheless, TDMA may exhaust too much time to accomplish a round of cooperative relaying. This problem becomes even worse, with the increment of the number of sources or relays, because a relatively large transmission delay may be resulted in. For the sake of achieving the same diversity order of TDMA but with less time slots, a new cooperative relaying scheme, i.e., space-time network coding (STNC), was represented in [6]. In a cooperative relaying system with $M$ sources transmitting their information to a common destination via $K$ relays, STNC is able to achieve the full diversity order ($K+1$) for each source with only ($M+K$) time slots, where if traditional TDMA is employed, ($M+MK$) time slots have to be consumed.

Owing to its potential capacities, STNC has attracted considerable attention in the past few years [7-11]. In [7-10], the outage and symbol error rate (SER) performance of STNC with amplify-and-forward (AF) or decode-and-forward (DF) relay operation was studied. Later, to further enhance the system performance and utilize the broadcast benefit of wireless links, the authors in [11] presented a new STNC-based cooperative relaying scheme, i.e., STNC with overhearing relays (STNC-OR), by combining the signals overheard from the previous relays with those received from the sources, where the outage and SER performance of STNC-OR with DF relaying operation was also analysed.

Although the outage probability of STNC with AF relay operation was studied in [10], the outage performance of STNC with overhearing AF relays has not been studied yet. Considering that each relay between the source and the destination may also employ diversity to improve the system performance with overhearing AF relays, this paper therefore aims to investigate the outage performance STNC with overhearing AF relays.

The contributions of the paper are summarized as follows. Firstly, we present a cooperative communication system with overhearing AF relays based on STNC (STNC-OHAF). Secondly, we theoretically analyse the instantaneous end-to-end SNR expressions for it. Thirdly, we study the outage probability of STNC-OHAF over independent but not necessarily identically distributed (i.n.i.d.) Rayleigh fading channels and then derive an explicit expression for it. Finally, we present some numerical results, which validate our analysis and show that with overhearing relays, the outage performance of the system can be greatly improved. It also shows that there is a trade-off between system sum outage capacity and the transmitted number of symbols.

The rest of the paper is organized as follows. Section II describes the system model. Section III analyses the instantaneous end-to-end SNR and the outage probability. In Section IV, we provide some numerical results and finally summarize the paper with some conclusions in Section V.

Manuscript received April XX, 20XX; accepted April XX, 20XX.
This research was supported by the Open Research Fund of National Mobile Communications Research Laboratory, Southeast University (no. 2014D03) and the Fundamental Research Funds for the Central Universities (no. 2014JBM024).

## II. SYSTEM MODEL

Consider a cooperative communication system, as shown in Figure 1, which consists of one source, $K$ relays and one destination. The source, the $r$-th relay and the destination are denoted as S, $R_r$ and D, respectively, where $r=1, 2, ..., K$. S wants to transmit its information to D with the help of the $K$ relays. For a more general consideration, we assume that the direct link between S and D exists in the system. Half-duplex mode is employed so that each node cannot transmit and receive signals at the same time.

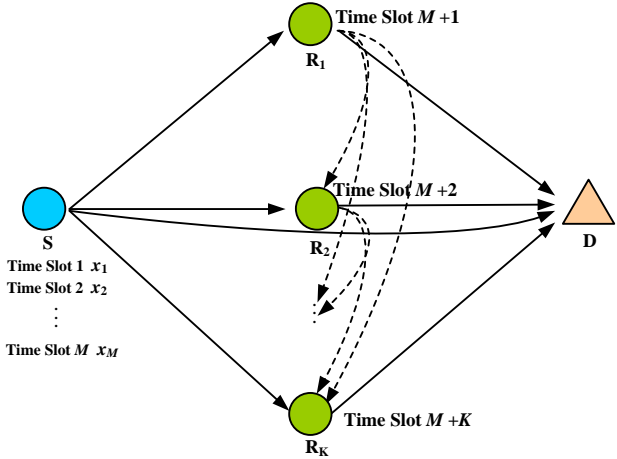

Fig.1. System model of STNC-OHAF, where the dotted arrows represent the overhearing links among the relays.

To facilitate the cooperative relaying transmission from S to D in such a system for each time period $T$, $T$ is divided into a certain amount of time slots with equal-interval $t$. In both STNC-AF and STNC-OHAF, $T$ is divided into $(M+K)$ slots, where the first $M$ time slots are assigned to the source phase and used for the source to broadcast its $M$ symbols one by one to the $K$ relays and D, and the rest $K$ time slots are assigned to the relay phase and used for the $K$ relays to forward their received signals to D one by one in a TDMA manner. Specifically, in the relay phase of STNC-AF, $R_r$ generates a linearly coded signal with its received signals from S and then amplify and forward the coded signal to D, while in STNC-OHAF, $R_r$ generates the linearly coded signal not only with the signals transmitted from S but also with the signals overheard from the previous relay $R_i$, where $i=1, 2, ..., r-1$.

The transmitted symbol in the $m$-th time slot from S is $x_m$, where $m=1, 2, ..., M$. All channels are modelled as i.n.i.d. quasi-static narrow-band Rayleigh fading channels with additive white Gaussian noise (AWGN). That is, the channel coefficients can be treated as constants over each transmission time period $T$ and independently change between two consecutive time periods. If a receiver receives at least two copies of the same transmitted signal from either the relays or the source, the receiver can combine them with maximum-ratio-combining (MRC) method to obtain a better performance.

Let $h_{u,v}$ represent the channel coefficient between terminals $u$ and $v$, where $v$ is the receiver (relays and the destination) and $u$ is the transmitter (source and relays). Therefore, $h_{u,v}$ is modelled as complex circular symmetric Gaussian random variables (RV) with mean equal to zero, i.e., $h_{u,v} \sim \mathcal{CN}(0, \sigma_{u,v}^2)$. Thus, $|h_{u,v}|^2$ is an exponentially distributed random variable with a parameter $\lambda_{u,v} = 1/\sigma_{u,v}^2$. The zero-mean and $N_0$-variance AWGN at receiver $v$ over link $u \to v$ is denoted by $n_{u,v}$, where $n_{u,v} \sim \mathcal{CN}(0, N_0)$. $\gamma_{u,v}$ and $\Gamma_{u,v}$ are used to denote the received SNR at $v$ over the direct link $u \to v$ between the two adjacent nodes and the total received SNR by using MRC method at $v$ over all possible paths from $u$ to $v$ (signals from the source $u$ and other relays $r$, $r<v$), respectively. The RV $\gamma_{u,v} = P_u |h_{u,v}|^2 / N_0$, is exponentially distributed with parameter $\zeta_{u,v} = 1/\bar{\gamma}_{u,v}$. $\bar{\gamma}_{u,v} = P_u \sigma_{u,v}^2 / N_0$ is the mean of $\gamma_{u,v}$ and $P_u$ is the power used by terminal $u$.

## III. PERFORMANCE ANALYSIS

In this section, we shall analyse the instantaneous end-to-end SNR and then derive the outage probability for STNC-OHAF.

### A. Instantaneous end-to-end SNR Analysis

In $m$-th time slot $t_m$ ($1 \leq m \leq M$), S broadcasts the signal $\sqrt{P_s} x_m c_m$, where $\sqrt{P_s}$ is the average transmit power at S and $c_m$ is the spreading code associated with $x_m$. With the perfect orthogonal assumption, $\langle c_m, c_m \rangle = 1$ and $\langle c_m, c_n \rangle = 0$, where $m \neq n$. Therefore, there is no interference between $x_m$ and $x_n$ if a matched-filter is adopted at the receiver to decode $x_m$. The received signal at $R_r$ is $y_{m,r} = \sqrt{P_s} h_{s,r} x_m c_m + n_{s,r}$.

For convenience, we define $\varphi_{s,r} \triangleq \sqrt{P_s} h_{s,r}^* / N_{s,r}$, which is the ratio of the conjugate of the signal component coefficient to the noise power and can be regarded as the MRC coefficient. Therefore $y_{m,r}$ can be expressed as

$$y_{m,r} = N_{s,r} \varphi_{s,r} x_m c_m + n_{s,r}. \quad (1)$$

In this case, after the source phase, $R_1$ receives $M$ signals from S and then generates a linearly combined signal, which can be expressed by

$$y_1^{(c)} = \sum_{m=1}^{M} \varphi_{s,1}(N_{s,1} \varphi_{s,1} x_m c_m + n_{s,1}). \quad (2)$$

Then, $R_1$ amplifies the coded symbol with an amplifier factor $\alpha_1$ and forwards $X_1 = \alpha_1 y_1^{(c)}$ to D. $\alpha_1$ is chosen such that the average power used by $R_1$ is equal to $P_1$, which is a predetermined desired value of power. The average transmit power of relay $R_1$ is

$$P_1 = \mathbb{E}[|X_1|^2] = \alpha_1^2 \sum_{m=1}^{M} \left( (N_{s,1}|\varphi_{s,1}|^2)^2 + \mathbb{E}[|\varphi_{s,1} n_{s,1}|^2] \right)$$
$$= \alpha_1^2 \sum_{m=1}^{M} ( \underbrace{(N_{s,1}|\varphi_{s,1}|^2)^2}_{A_1} + \underbrace{(N_{s,1}|\varphi_{s,1}|^2)}_{A_1} ) \quad (3)$$
$$= \alpha_1^2 \sum_{m=1}^{M} (|A_1|^2 + A_1) = \alpha_1^2 M(|A_1|^2 + A_1)$$

where $\mathbb{E}[\cdot]$ denotes the expectation operator. Thus, it can be deduced that $\alpha_1 = \sqrt{P_1}/\sqrt{M(|A_1|^2 + A_1)}$.

For $R_r$ ($r=2, ..., K$), due the broadcast nature, at the end of time slot $t_{m+r-1}$, it not only received the $M$ transmitted signals from S but also collected the transmitted signal $X_i$ from $R_i$, where $i=1, 2, ..., r-1$. For example, the collected signal at $R_r$ from $R_1$ can be given by

$$y_{1,r} = h_{1,r} X_1 + n_{1,r} = h_{1,r} \alpha_1 \sum_{m=1}^{M} \varphi_{s,1}(N_{s,1} \varphi_{s,1} x_m c_m + n_{s,1}) + n_{1,r}$$

$$= \sum_{m=1}^{M} \chi_{1,r}\varphi_{1,r}x_m c_m + \eta_{1,r}, \text{ for } r=2, ..., K. \quad (4)$$

where $\eta_{1,r} = n_{1,r} + Mh_{1,r}\alpha_1\varphi_{s,1}n_{s,1}$ which is the total received noise signal at $R_r$ from $R_1$. $\varphi_{1,r} = h_{1,r}\alpha_1 A_1 / \chi_{1,r}$, where $\chi_{1,r} = \mathbb{E}[|\eta_{1,r}|^2] = N_{1,r} + M |h_{1,r}|^2 \alpha_1^2 |\varphi_{s,1}|^2 N_{s,1}$. Therefore, at the end of time slot $t_{m+r-1}$, $R_r$ collects $r$ signal copies associated with $x_m$, the first copy is received in the time slot $t_m$ from S, i.e., $y_{m,r}$ and the $k$-th ($2 \le k \le r$) copy is in the time slot $t_{M+k-1}$ from $R_{k-1}$, which can be extracted from $y_{k-1,r}$ by $\langle y_{k-1,r}, c_m \rangle$ and the extracted signal copy associated with $x_m$ is $y_{k-1,r}^{(m)} = \chi_{k-1,r}\varphi_{k-1,r}x_m + \langle \eta_{k-1,r}, c_m \rangle$.

Then, $R_r$ processes each copy by multiplying a MRC coefficient with it and then combines the $r$ processed signal copies for $x_m$ as

$$y_r^{(m)} = \varphi_{s,r}(N_{s,r}\varphi_{s,r}x_m c_m + n_{s,r}) + \sum_{i=1}^{r-1}\varphi_{i,r}(\eta_{i,r} + \chi_{i,r}\varphi_{i,r}x_m c_m). \quad (5)$$

After this, in the time slot $t_{m+r}$, $R_r$ generates a linearly coded signal $y_r^{(c)}$ with the $M$ combined signals of $x_1, x_2, ..., x_M$ as $y_r^{(c)} = \sum_{m=1}^{M} y_r^{(m)}$ and amplifies it with a factor $\alpha_r$, producing a newly transmitted symbol $X_r = \alpha_r y_r^{(c)}$, where $\alpha_r = \sqrt{P_r}/\sqrt{M(|A_r|^2 + A_r)}$, $A_r = N_{s,r}|\varphi_{s,r}|^2 + \sum_{i=1}^{r-1}\chi_{i,r}|\varphi_{i,r}|^2$. Then, $R_r$ broadcast $X_r$ to D and all its succeeding relay $R_j$ ($j=r+1, ..., K$). Thus, the received signals at $R_j$ can be given by (6), where $\varphi_{r,j}$ and $P_r$ can be seen in (7) and (8).

Similarly, at the end of $T$, D collects $K+1$ signal copies associated with $x_m$ from S and the $K$ relays. Also, with the similar signal extraction operation as shown in (5), the total received signal of $x_m$ can be given by

$$y_d^{(m)} = \underbrace{(N_{s,d}|\varphi_{s,d}|^2 + \sum_{r=1}^{K}\chi_{r,d}|\varphi_{r,d}|^2)}_{A_d}x_m c_m + \underbrace{(\varphi_{s,d}n_{s,d} + \sum_{r=1}^{K}\varphi_{r,d}\eta_{r,d})}_{\eta_d}$$

(9)

**Lemma 1.** The accurate approximate instantaneous end-to-end SNR associated with $x_m$ at the D of STNC-OHAF is

$$\Gamma_{s,d}^{(m)} \approx \gamma_{s,d} + \frac{1}{M}\sum_{r=1}^{K}\frac{A_r\gamma_{r,d}}{A_r + \gamma_{r,d} + 1}, \quad (10)$$

where

$$A_r = N_{s,r}|\varphi_{s,r}|^2 + \sum_{i=1}^{r-1}\chi_{i,r}|\varphi_{i,r}|^2 = \gamma_{s,r} + \frac{1}{M}\sum_{i=1}^{r-1}\frac{A_i\gamma_{i,r}}{A_i + \gamma_{i,r} + 1},$$

and $\gamma_{u,v} = p_u|h_{u,v}|^2/N_{u,v}$, ($u=s,1,...,K$; $v=u+1,...,K,d$), which is the instantaneous SNR for the channel from terminal $u$ to $v$.

**Proof:** Since $\mathbb{E}[|\eta_d|^2] \approx N_{s,d}|\varphi_{s,d}|^2 + \sum_{r=1}^{K}\chi_{r,d}|\varphi_{r,d}|^2$, according to (9), the instantaneous end-to-end SNR associated with $x_m$ can be given by

$$\Gamma_{s,d}^{(m)} = \frac{A_d^2}{\mathbb{E}[|\eta_d|^2]} \approx N_{s,d}|\varphi_{s,d}|^2 + \sum_{r=1}^{K}\chi_{r,d}|\varphi_{r,d}|^2.$$

Further, it can be deduced that

$$N_{s,d}|\varphi_{s,d}|^2 = N_{s,d}(\sqrt{P_s}h_{s,d}/N_{s,d})^2 = P_s|h_{s,d}|^2/N_{s,d} = \gamma_{s,d}.$$

For $R_1$, we have that

$$\chi_{1,d}|\varphi_{1,d}|^2 = \frac{(h_{1,d}\alpha_1 N_{s,1}|\varphi_{s,1}|^2)^2}{M|h_{1,d}|^2\alpha_1^2|\varphi_{s,1}|^2 N_{s,1} + N_{1,d}}$$

$$= \frac{N_{s,1}|\varphi_{s,1}|^2|h_{1,d}|^2|\varphi_{s,1}|^2 p_1 N_{s,1}}{M|h_{1,d}|^2 p_1|\varphi_{s,1}|^2 N_{s,1} + N_{1,d}MN_{s,1}|\varphi_{s,1}|^2(N_{s,1}|\varphi_{s,1}|^2 + 1)}$$

$$= \frac{N_{s,1}|\varphi_{s,1}|^2|h_{1,d}|^2 p_1/N_{1,d}}{M|h_{1,d}|^2 p_1/N_{1,d} + M(N_{s,1}|\varphi_{s,1}|^2 + 1)} = \frac{1}{M}\frac{A_1\gamma_{1,d}}{A_1 + \gamma_{1,d} + 1}.$$

For $R_r$ ($r=2, ..., K$), we have that

$$\chi_{r,d}|\varphi_{r,d}|^2 = \chi_{r,d}(\alpha_r A_r h_{r,d}/\chi_{r,d})^2 = \alpha_r^2 A_r^2|h_{r,d}|^2/\chi_{r,d}$$

$$= \frac{(A_r+1)\alpha_r^2 A_r^2|h_{r,d}|^2}{|h_{r,d}|^2 p_r + (A_r+1)N_{r,d}} = \frac{1}{M}\frac{A_r|h_{r,d}|^2 p_r/N_{r,d}}{|h_{r,d}|^2 p_r/N_{r,d} + A_r + 1}$$

$$= \frac{1}{M}\frac{A_r\gamma_{r,d}}{A_r + \gamma_{r,d} + 1}.$$ Thus, Lemma 1 is proved. ∎

### B. Outage Probability Analysis

**Theorem 1.** The outage probability $P_{out}$ of STNC-OHAF can be approximated by

$$P_{out} \approx [(K+1)!]^{-1}\left(2^{(M+K)R} - 1\right)^{K+1}\zeta_{s,d}\left(\zeta_{1,d} + \zeta_{s,1}\right)\prod_{r=2}^{K}\zeta_{r,d}. \quad (11)$$

**Proof:** Let R be the required transmission rate of S. It is assumed that if $\mathcal{I}_{s,d} = \log_2(1 + \Gamma_{s,d}^{(m)})/(M+K) < R$, the outage event occurs. Thus,

$$P_{out} = \mathbb{P}[\mathcal{I}_{s,d} < R] = \mathbb{P}[\Gamma_{s,d}^{(m)} < 2^{(M+K)R} - 1] = \int_0^{2^{(M+K)R}-1}\mathbb{F}\left(\Gamma_{s,d}^{(m)} = \gamma\right)d\gamma$$

where $\mathbb{F}(\square)$ and $\mathbb{P}[\square]$ denote the probability density function (PDF) and the cumulative distribution function (CDF) respectively. By using the method developed in [10], $P_{out}$ can be approximated by the first term of $\Gamma_{s,d}^{(m)}/\Gamma$'s MacLaurin series, where $\Gamma$ denotes the average SNR.

$$P_{out} \approx \frac{\left(2^{(M+K)R} - 1\right)^{K+1}}{(K+1)!}\frac{\partial^K \mathbb{F}\left(\Gamma_{s,d}^{(m)} = \gamma\right)}{\partial \gamma^K}(0).$$

By some algebraic manipulations and the Lemma 2 in [10], Theorem 1 is proved. ∎

## IV. RESULTS AND DISCUSSION

In this section, we present some numerical results to discuss the performance of the proposed STNC-OHAF. For comparison, the performance of STNC-AF in [10] and traditional pure TDMA with overhearing relays (TDMA-OH) are also simulated.

In the numerical experiments, a network model as shown in Figure 1 is considered. We assume the unit noise variance, i.e., $N_0=1$, and the normalized time period, i.e., $T=1$. Perfect orthogonal spreading codes are used, and $P_s = P_r = P_{tot}/(K+M)$, where $P_{tot}$ is the total available power of the system. With these assumptions, the performance is expected to be the same for all $x_m$ and therefore the performance associated with $x_1$ can be presented as representative to compare discuss the outage performances of different schemes in terms of outage probabilities. Moreover, we define the average system SNR=$P/N_0$ and the values of $\sigma_{u,v}^2$ are randomly selected within [0.1, 25]. Note that these parameter settings will not change in our following numerical examples and simulations unless otherwise specified.

Figure 2 plots the outage probability of the three schemes versus the transmit SNR. Firstly, it can be seen that the numerical results of STNC-OHAF match the simulation ones well, especially in the high SNR region, which demonstrated

$$y_{r,j} = h_{r,j}X_r + n_{r,j} = h_{r,j}\alpha_r \sum_{m=1}^{M}((N_{s,r}|\varphi_{s,r}|^2 + \sum_{i=1}^{r-1}\chi_{i,r}|\varphi_{i,r}|^2)x_m c_m + \varphi_{s,r}n_{s,r} + \sum_{i=1}^{r-1}\varphi_{i,r}\eta_{i,r}) + n_{r,j} \qquad (6)$$

$$= \sum_{m=1}^{M} h_{r,j}\alpha_r A_r x_m c_m + M h_{r,j}\alpha_r(\varphi_{s,r}n_{s,r} + \sum_{i=1}^{r-1}\varphi_{i,r}\eta_{i,r}) + n_{r,j} = \sum_{m=1}^{M}\chi_{r,j}\varphi_{r,j}x_m c_m + \eta_{r,j},$$

$$\varphi_{r,j} = h_{r,j}\alpha_r A_r / \chi_{r,j}, \eta_{r,j} = M h_{r,j}\alpha_r(\varphi_{s,r}n_{s,r} + \sum_{i=1}^{r-1}\varphi_{i,r}\eta_{i,r}) + n_{r,j}, \chi_{r,j} = \mathbb{E}[|\eta_{r,j}|^2], \qquad (7)$$

$$P_r = \mathbb{E}[|y_r^{(t)}|^2] = \alpha_r^2 \sum_{m=1}^{M}\left((N_{s,r}|\varphi_{s,r}|^2 + \sum_{i=1}^{r-1}\chi_{i,r}|\varphi_{i,r}|^2)^2 + \mathbb{E}[|\varphi_{s,r}n_{s,r} + \sum_{i=1}^{r-1}\varphi_{i,r}\eta_{i,r}|^2]\right) \qquad (8)$$

$$= \alpha_r^2 \sum_{m=1}^{M}\left((N_{s,r}|\varphi_{s,r}|^2 + \sum_{i=1}^{r-1}\chi_{i,r}|\varphi_{i,r}|^2)^2 + N_{s,r}|\varphi_{s,r}|^2 + \sum_{i=1}^{r-1}\chi_{i,r}|\varphi_{i,r}|^2\right) = \alpha_r^2 M(|A_r|^2 + A_r)$$

the correctness of our theoretical analysis. Secondly, for the same $K$, the three schemes achieve the same diversity order. Thirdly, STNC-OHAF always achieves the lowest outage probabilities among the three ones, which implies that by introducing the overhearing among relays, the system performance can be greatly enhanced.

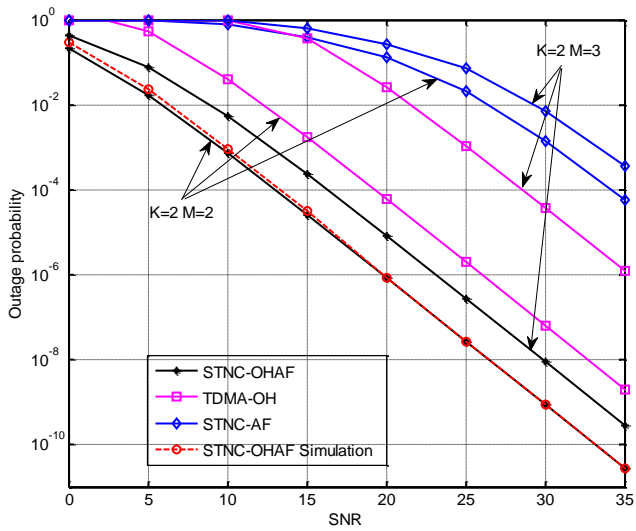

Fig.2. Comparison of outage probability versus transmit SNR.

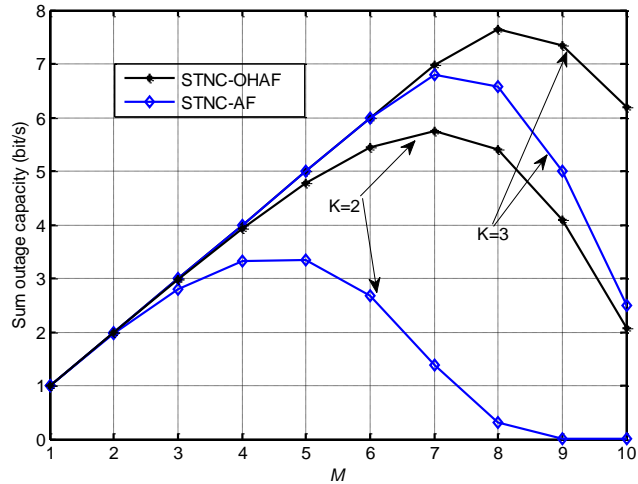

Fig.3. Comparison of sum outage capacity versus $M$.

Moreover, in order to discuss the effect of the number of transmitted symbols $M$ on the system performance, Figure 3 shows the sum outage capacity of the system versus $M$. The sum outage capacity $C_{soc}$ is defined by

$$C_{soc} \triangleq \sum_{m=1}^{M}(1-P_{out,s})\frac{1}{M+K}B\log_2(1+\gamma_{min}) = \sum_{m=1}^{M}(1-P_{out,s})BR$$

where $\gamma_{min}$ is the minimum received SNR, The bandwidth $B$ and transmission rate threshold $R$ are normalized to unity. In the simulations, $K=2, 3$ and transmit SNR is 25 dB. It shows that there is an optimal value of $M$ in terms of maximizing the sum outage capacity for both STNC-AF and STNC-OHAF schemes. When $M$ is small, the systems benefit from spatial multiplexing gain whereas when $M$ becomes relatively large, for a given $T$, the decreasing time slot internal limits the system performance, which means that increasing $M$ cannot always improve the system outage capacity. Besides, it also shows that STNC-OHAF always achieves much higher sum outage capacity than STNC-AF.

## V. CONCLUSIONS

This paper presented a new cooperative relaying, STNC-OHAF, by allowing each relay to collect transmitted signal from the source and its previous relays. We derived the expressions of the SNR and the outage probability for the proposed STNC-OHAF over Rayleigh fading channels. It is shown that with overhearing relays, the outage performance of the system can be greatly improved and there is a trade-off between system sum outage capacity and the transmitted number of symbols.